\documentclass[prb,twocolumn,showpacs]{revtex4}
\usepackage{graphicx}

\begin{document}

\title{
   Spin phase diagram of the $\nu_e=4/11$ composite fermion liquid}

\author{
   Arkadiusz W\'ojs,$^{1,2}$
   George Simion,$^2$ and
   John J. Quinn$^2$}

\affiliation{
   \mbox{
   $^1$Institute of Physics, Wroc{\l}aw University of Technology,
       Wybrze\.ze Wyspia\'nskiego 27, 50-370 Wroc{\l}aw, Poland}\\
   $^2$Department of Physics, University of Tennessee, 
       Knoxville, TN 37996, USA}

\begin{abstract}
Spin polarization of the ``second generation'' $\nu_e=4/11$ fractional 
quantum Hall state (corresponding to an incompressible liquid in a 
one-third-filled composite fermion Landau level) is studied by exact 
diagonalization.
Spin phase diagram is determined for GaAs structures of different 
width and electron concentration.
Transition between the polarized and partially unpolarized states 
with distinct composite fermion correlations is predicted for realistic 
parameters.
\end{abstract}
\maketitle

\section{Introduction}

There has been considerable speculation about the nature 
of ``second-generation'' incompressible quantum liquid (IQL) 
states observed recently by Pan {\em et al.}\cite{Pan03}
Their incompressibility depends on spin and charge dynamics 
of the fractionally charged Laughlin quasiparticles (QPs).
\cite{Laughlin83} 

Pan's experiment employed the fractional quantum Hall (FQH) 
effect,\cite{Tsui82} a non-perturbative interaction many-body 
phenomenon, dependent on magnetic quantization of the 
two-dimensional single-electron energy spectrum into massively 
degenerate Landau levels (LLs).\cite{Yoshioka02}
It coincides with the formation of electron IQLs and thus 
occurs at the particular fractional values of the LL filling 
factor, defined as $\nu_e=2\pi\varrho_e\lambda^2$ (where 
$\varrho_e$ is the electron concentration and $\lambda=
(\hbar c/eB)^{1/2}$ is the magnetic length).

The emergence of IQLs is explained as follows by the composite 
fermion (CF) theory.\cite{Jain89} 
Electrons partially filling the lowest LL are said to capture $2p$ 
magnetic flux quanta $\phi_0=hc/e$ and become (weakly interacting) 
CFs moving in a reduced effective magnetic field, corresponding to 
a higher effective CF filling factor $\nu_{\rm CF}$.
The most prominent IQL sequence at $\nu_e=s(2ps\pm1)^{-1}$ (with 
$s$ and $p$ being a pair of integers) corresponds to $\nu_{\rm CF}=s$, 
i.e., to the integral quantum Hall effect of the CFs.

However, not all IQLs found in the lowest LL can be explained in 
this way.
Recently, Pan {\em et al.}\cite{Pan03} observed the FQH effect 
at $\nu_e={4\over11}$, corresponding to $\nu_{\rm CF}={4\over3}$, 
i.e., to a partal filling of a CF-LL.
This discovery demonstrated that CFs, like electrons, can form IQLs.
The origin of incompressibility of Pan's correlated CF liquid (also 
called a ``second-generation'' FQH state) has been vigorously studied 
for the last three years.
\cite{Smet03,Chang04,Lopez04,qepair,Goerbig04,clusters,Quinn06}
However, some of even most fundamental questions remain controversial.

The subject of this paper is polarization of the $\nu_e={4\over11}$ 
state.
It is largely motivated by the wealth of theory of spin dynamics in 
the ``first-generation'' FQH states.\cite{Halperin83,Chakraborty84,%
Rezayi87,Wu93,Apalkov01,Kamilla96,MacDonald98,sky}
However, our main goal is to extend the work of Chang {\em et al.}
\cite{Chang03} and directly address Pan's experimental results in tilted 
magnetic fields\cite{Pan03} which indicated ferromagnetic order.
In the CF picture, this corresponds to a completely filled lowest CF-LL 
(0$\uparrow$) and a ${1\over3}$-filled first excited CF-LL with the same 
spin (1$\uparrow$).
Since the Laughlin $\nu={1\over3}$ state in CF-LL$_1$ was earlier ruled 
out\cite{hierarchy} based on the form of short-range CF--CF interaction 
pseudopotential, the explanation for the observed incompressibility must 
be different.
This distinction makes the polarized $\nu_e={4\over11}$ state an object 
of intense investigation.\cite{Mandal02}
Although several ideas were formulated (e.g., CF pairing
\cite{qepair,clusters}), neither an analytic CF wavefunction nor an 
intuitive understanding for the incompressibility has been reached.
A partially unpolarized state was also proposed,\cite{Park00} 
with the $\nu={1\over3}$ filling of the lowest CF-LL with reversed 
spin (0$\downarrow$).
In contrast to the polarized state and due to a different form\cite{qer} 
of CF--CF interaction in CF-LL$_0$, it is expected to be a Laughlin CF 
liquid.
However, this state has not yet been observed in experiment.

Let us summarize this remarkable situation as follows:
The polarized state has been observed but it is not well understood,
and the unpolarized state has not been observed but it appears to be 
much easier to understand.
In this paper we calculate the single-particle and correlation energies 
in these two competing CF states, depending on the experimenetally 
controlled parameters (electron layer width, concentration, and magnetic 
field).
The main result is the spin phase diagram, from which we predict a spin
transition at $\nu_e={4\over11}$, induced e.g.\ by an additional electric
field narrowing the electron layer.
Suggested experimental demonstration of this transition would shine more 
light on the role played by spin of correlated CFs.

\section{Numerical model}

The calculations were done in Haldane's spherical geometry,
\cite{Haldane83} convenient for the numerical studies of 
incompressible quantum liquids with short-range correlations.
To model an extended (planar) 2D system of interacting particles
filling a fraction $\nu$ of a degenerate LL, their finite number 
$N$ is considered within a shell of appropriate angular momentum 
$l$ and degeneracy $g=2l+1$ (containing states with different 
angular momentum projections, $|m|\le l$).
The assignment of the filling factor $\nu$ to a finite system
$(N,g)$ is not trivial. 
It requires identifying dependence $g=\nu^{-1}N+\gamma$ which 
defines a series of finite systems representing an infinite state 
$\nu$ (here, the ``shift'' $\gamma$ is independent of $N$ but 
it depends on the form of correlations, i.e., in particular 
on $\nu$).

In the original formulation,\cite{Haldane83,Fano86} these $l$-shells 
represent LLs of a charged particle confined to a surface of 
a sphere of radius $R$, with the normal magnetic field $B$ produced 
by a Dirac monopole of strength $2Q=4\pi R^2 B/\phi_0$.
Specifically, the $n$th LL on a plane (called LL$_n$; with $n\ge0$) 
corresponds to the shell of $l=Q+n$ on a sphere.

Here, we do not use the particular form of the $\left|Q;n,m\right>$ 
wavefunctions, but take advantage of the fact that the symmetry of
angular momentum eigenstates $\left|l,m\right>$ under 2D rotations 
mimics the symmetry of the planar eigenstates under 2D (magnetic)
translations.
Thus, the interaction matrix elements are guaranteed to obey general 
rules for a scalar operator in the basis of spherical harmonics, 
but the particular values are put into the model ``by hand,'' so as 
to describe the actual interaction among the considered particles 
(on the plane).
This is done by specifying Haldane pseudopotential,\cite{Haldane87} 
defined as interaction energy $V$ as a function of relative angular 
momentum $\mathcal{R}$.
On a sphere relative and total pair angular momenta are related
by $\mathcal{R}+L=2l$, and the matrix elements $\left<l_1,m_1;
l_2,m_2|V|l_3,m_3;l_4,m_4\right>$ are connected with $V(L)$ through
the Clebsch-Gordan coefficients.

The many-body interaction Hamiltonian is diagonalized numerically 
in the configuration-interaction basis, using a Lanczos algorithm.
The energy levels $E$ are determined separately for each subspace 
of the total spin $S$ and angular momentum $L$.

\section{Single-quasielectron energies}

In the mean-field CF transformation, the liquid of Laughlin 
correlated electrons at filling factor $\nu_e={4\over11}$ 
is converted to the system of CFs with an effective filling 
factor $\nu_{\rm CF}={4\over3}$.
Thus, the low-energy dynamics of $N_e$ electrons with Coulomb
interaction in the lowest LL can be mapped onto that of $\sim
{3\over4}N_e$ CFs completely and rigidly filling the lowest CF-LL 
(0$\uparrow$) and the excess of $N\approx{1\over4}N_e$ 
CFs in the $\nu={1\over3}$ filled next lowest CF-LL (either 
1$\uparrow$ or 0$\downarrow$, depending on the relative 
magnitude of electron Zeeman energy $E_{\rm Z}$ and the 
effective CF cyclotron gap $\propto e^2/\lambda$).
Each CF in the partially filled 1$\uparrow$ or 0$\downarrow$
LL represents a ``normal''\cite{Laughlin83} or ``reversed-spin''
\cite{Rezayi87} quasielectron (QE or QE$_{\rm R}$) of the 
underlying incompressible Laughlin liquid, respectively.

The Coulomb energies $\varepsilon_{\rm QE}$ and $\varepsilon_{\rm 
QER}$ of these two QPs can be extracted\cite{qer,Fano86} from exact 
diagonalization of finite systems of $N_e$ electrons in the lowest 
LL with the appropriate degeneracy $g$.
The Laughlin ground state occurs at $g=3N_e-2\equiv g_{\rm L}$; 
it is non-degenerate ($L=0$) and spin-polarized ($S={1\over2}N_e$).
A single QE or QE$_{\rm R}$ appears in the Laughlin liquid in 
the lowest states at $g=g_{\rm L}-1$ and either $S={1\over2}N_e$ 
or ${1\over2}N_e-1$, respectively.
The QE and QE$_{\rm R}$ energies $\varepsilon$ (defined relative 
to the underlying Laughlin liquid) are obtained from the comparison 
of the ($N_e$-electron) energies at $g=g_{\rm L}$ and and $g_{\rm L}-1$.

\begin{figure}
\includegraphics[width=3.4in]{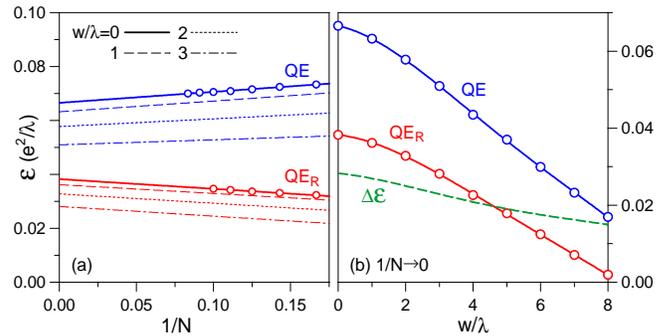}
\caption{(color online)
   Dependence of the quasielectron (QE) and reversed-spin 
   quasielectron (QE$_{\rm R}$) energies $\varepsilon$ on:
   (a) the inverse electron number $N^{-1}$ in a finite-size 
   calculation, and 
   (b) the electron layer width $w$.
   $\lambda$ is the magnetic length.}
\label{fig1}
\end{figure}

The numerical procedure and the result for an ideal 2D electron 
layer were presented earlier.\cite{qer,Fano86} 
In Fig.~\ref{fig1} we compare the QE/QE$_{\rm R}$ energies 
calculated for quasi-2D layers of finite width $w$.
Here, $w$ is the effective width of the electron wavefunction 
in the normal ($z$) direction, approximated by $\chi(z)\propto
\cos(z\pi/w)$.
It is slightly larger than the quantum well width $W$; e.g., 
for symmetric GaAs/Al$_{0.35}$Ga$_{0.65}$As wells, $w\approx 
W+3.3$~nm over a wide range of $W\ge10$~nm.
The regular dependence on system size in Fig.~\ref{fig1}(a)
allows reliable extrapolation of $\varepsilon$ to $N^{-1}
\rightarrow0$ (planar geometry).
From the comparison of $\varepsilon_{\rm QE}(w)$ and 
$\varepsilon_{\rm QER}(w)$ in Fig.~\ref{fig1}(b) it is clear 
that their difference $\Delta\varepsilon$ is less sensitive 
to the width than any of the $\varepsilon$'s.
To put the shown width range in some perspective, let us note
that a (fairly narrow) $W=12$~nm well in a (fairly low) field 
$B=10$~T corresponds to $w/\lambda=1.9$ and $\Delta\varepsilon
(w)/\Delta\varepsilon(0)=0.9$, justifying the 2D approximation.
On the other hand, a wide $W=40$~nm well in a high field 
$B=23$~T gives $w/\lambda=8.1$ and $\Delta\varepsilon(w)/
\Delta\varepsilon(0)=0.5$, i.e., a significant width effect.

\section{Quasielectron interactions}

The weak effective CF--CF interactions are known with some accuracy
from earlier studies.\cite{hierarchy,clusters,qer,Lee01,Sitko97}
At least at sufficiently low CF fillings factors $\nu\le{1\over3}$, 
they can be well approximated by fixed Haldane pseudopotentials 
(independent of the CF-LL filling or spin polarization).
The short-range QE--QE, QE$_{\rm R}$--QE$_{\rm R}$, and QE--QE$_{\rm R}$ 
pseudopotentials can be obtained from finite-size diagonalization for
$N_e$ electrons with up to two revesed spins ($S={1\over2}N_e-2$)
at $g=g_{\rm L}-2$.

The result is a reliable account of the relative values $\Delta 
V_{\mathcal{R},\mathcal{R}'}=V(\mathcal{R})-V(\mathcal{R}')$ at small 
neighboring $\mathcal{R}$ and $\mathcal{R}'$, but the absolute values 
are not estimated very accurately.
Fortunately, since vertical correction of $V(\mathcal{R})$ by a constant 
does not affect the many-CF wavefunctions and only rigidly shifts the 
entire energy spectrum, a few leading values of $\Delta V$ completely 
determine the (short-range) CF correlations at a given $\nu$.
Therefore, the knowledge of those few approximate values of $\Delta 
V_{\rm QER}$ and $\Delta V_{\rm QE}$ was sufficient to establish that:
(i) the QE$_{\rm R}$'s form a Laughlin $\nu={1\over3}$ liquid
\cite{Chang03,Park00,qer} which in finite $N$-QE$_{\rm R}$ systems 
on a sphere occurs at $g=3N-2$, and (ii) in contrast, the QEs form 
a different (probably paired) state\cite{qepair,clusters} at the same 
$\nu={1\over3}$, which on a sphere occurs at $g=3N-6$.

However, the relative strength of QE--QE and QE$_{\rm R}$--QE$_{\rm R}$ 
pseudopotentials $V_{\rm QER}$ and $V_{\rm QE}$ must also be known 
(in addition to $\Delta V$) to compare the energies of many-QE$_{\rm R}$ 
and many-QE states (i.e., of the spin-polarized and unpolarized electron 
states at $\nu_e={4\over11}$).
The absolute values of $V_{\rm QER}$ and $V_{\rm QE}$ can be obtained 
by matching\cite{Lee01} the short-range behavior from exact 
diagonalization of small systems with the long-range behavior predicted 
for a pair of charges $q=-{1\over3}e$.
Specifically, the short-range part of $V_{\rm QER}(\mathcal{R})$, which 
describes a pair of CFs in the 1$\downarrow$ CF-LL, is shifted to match 
$\eta V_0(\mathcal{R})$, the electron pseudopotential in the lowest LL 
rescaled by $\eta\equiv(q^2\lambda_q^{-1})/(e^2\lambda_e^{-1})=(q/e)^{5/2}$.
Similarly, the short-range part of $V_{\rm QE}(\mathcal{R})$, related to
the 1$\uparrow$ CF-LL, is shifted to match $\eta V_1(\mathcal{R})$.

\begin{figure}
\includegraphics[width=3.4in]{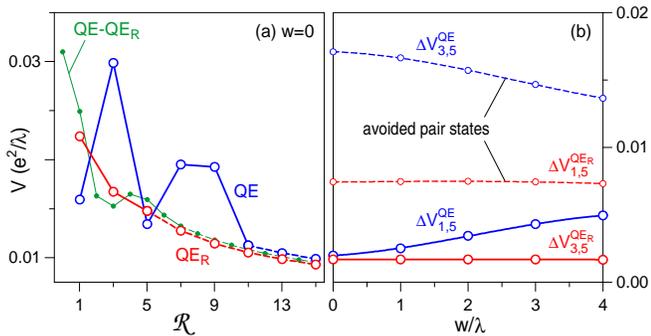}
\caption{(color online)
   (a) Haldane pseudopotentials (pair interaction energy $V$ 
   as a function of relative angular momentum $\mathcal{R}$)
   for quasielectrons (QE) and reversed-spin quasielectrons 
   (QE$_{\rm R}$) in an ideal 2D ($w=0$) electron layer.
   (b) Dependence of pseudopotential increments
   $\Delta V_{\mathcal{R}\mathcal{R}'}=V(\mathcal{R})-
   V(\mathcal{R}')$ on the electron layer width $w$.
   $\lambda$ is the magnetic length.}
\label{fig2}
\end{figure}

The result in Fig.~\ref{fig2}(a) for an ideal 2D layer was reported 
earlier;\cite{clusters} in Fig.~\ref{fig2}(b) the width 
dependence of the leading parameters $\Delta V$ has been plotted.
It is noteworthy that $V_{\rm QE}$ is much more sensitive to the electron 
layer width $w$ than $V_{\rm QER}$. 
This is explained by stronger oscillations in $V_{\rm QE}(\mathcal{R})$ 
at $w=0$, which tend to weaken in wider wells (when the characteristic 
in-plane distances decrease relative to $w$).
The curves for $V_{\rm QER}(1)$ and $V_{\rm QE}(3)$ have been drawn 
with dashed lines, since the QE$_{\rm R}$--QE$_{\rm R}$ and QE--QE 
pair states associated with these dominant pseudopotential parameters
will be avoided\cite{qepair} in the unpolarized and polarized $\nu=
{1\over3}$ CF ground states, respectively.

\section{Correlation energies of quasielectron liquids}

\begin{figure}
\includegraphics[width=3.4in]{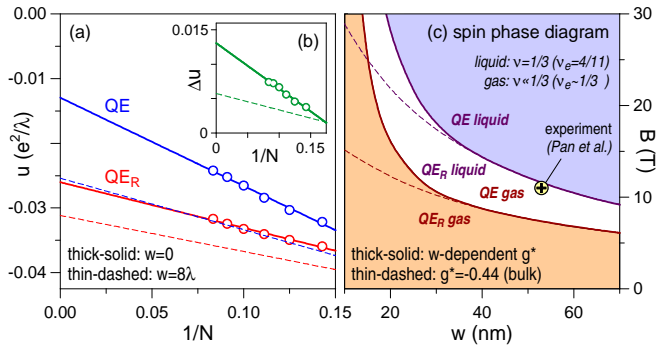}
\caption{(color online)
   (a) Correlation energy $u$ in the $\nu=1/3$ incompressible liquid 
   of quasielectrons (QE) or reversed-spin quasielectrons (QE$_{\rm R}$) 
   as a function of their inverse number $N^{-1}$, for two different 
   widths $w$ of the quasi-2D electron layer ($\lambda$ is the magnetic 
   length).
   (b) Difference $\Delta u=u_{\rm QE}-u_{\rm QER}$ as a function of 
   $N^{-1}$.
   (c) Phase diagram (critical layer width $w$ vs magnetic field $B$) 
   for the QE--QE$_{\rm R}$ spin transition at $\nu=1/3$ (i.e., 
   $\nu_e=4/11$), assuming effective electron $g$-factor for GaAs.
   Dashed line is for uncorrelated QEs or QE$_{\rm R}$'s (e.g., at 
   $\nu\ll1/3$).
   The experimental point taken after Pan {\em et al.}\cite{Pan03}}
\label{fig3}
\end{figure}

As mentioned above, due to the strong QE$_{\rm R}$--QE$_{\rm R}$ 
repulsion at short range ($\mathcal{R}=1$), the QE$_{\rm R}$'s form 
a Laughlin $\nu={1\over3}$ state similarly to the electrons at $\nu_e
={1\over3}$.
The corresponding series of non-degenerate $N$-QE$_{\rm R}$ ground 
states on a sphere occurs at the Lauglin sequence of $g=3N-2$.
In Fig.~\ref{fig3}(a) we plot the size dependence of their correlation 
energy $u$ (per particle), defined as 
\begin{equation}
   u={E+U_{\rm bckg}\over N}\,\zeta.
\label{eq1}
\end{equation}
Here, $E$ is the interaction energy of the ground state of $N$ 
QE$_{\rm R}$'s, $U_{\rm bckg}=-(Nq)^2/2R$ is a correction due to 
interaction with the charge-compensating background (with the
sphere radius $R=\lambda\sqrt{Q}$ taken for $2Q+1=g$, in analogy
to the relation for electrons in the lowest LL).
Factor $\zeta=\sqrt{Q(Q-1)^{-1}}$ is used to rescale the energy 
unit $e^2/\lambda=\sqrt{Q}\,e^2/R$ from that corresponding to 
$g_{\rm QER}=3N-2$ to that of an average $\bar{g}={1\over2}
(g_{\rm QER}+g_{\rm QE})=3N-4$, to allow for a later 
comparison of $u$ calculated for QE$_{\rm R}$'s and QE's at 
different $g$'s (and thus, at different magnetic lengths $\lambda$ 
corresponding to the same area $4\pi R^2$).

The correlation energies $u$ were calculated for $N\le12$, and 
extrapolated to $N^{-1}\rightarrow0$ to eliminate the finite-size 
effects.
Neither the particular form of $U_{\rm bckg}$ (i.e., the assumption 
of $g=2Q+1$ for the relation between $R$ and $\lambda$) nor the 
rescaling by $\zeta$ directly affect the extrapolated value 
(they only affect the size dependence, and thus the accuracy 
of extrapolation).
For an ideal 2D system, the result of extrapolation is 
$u_{\rm QER}=-0.026\,e^2/\lambda=-0.405\eta\,e^2/\lambda$.
This value is very close to $\eta u_0$, where $u_0=-0.412\,
e^2/\lambda$ describes the Laughlin state of electrons in LL$_0$.
Good agreement confirms not only Laughlin correlations among 
the QE$_{\rm R}$'s (which are guaranteed by the form of 
$\Delta V_{\rm QER}$ and can also be verified directly 
by the analysis of pair amplitudes) but, more importantly,
the accurate estimate of the absolute values of 
$V_{\rm QER}(\mathcal{R})$ drawn in Fig.~\ref{fig2}(a) and used 
in the $N$-QE$_{\rm R}$ calculation.

Let us turn to the QEs.
The dominant QE--QE repulsion at $\mathcal{R}=3$ causes the QEs 
to form pairs\cite{clusters} rather than a Laughlin state at $\nu=
{1\over3}$ (although the exact wavefunction of this incompressible 
state is still unknown).
The corresponding series of non-degenerate $N$-QE ground states 
on a sphere was identified\cite{qepair} at $g=3N-6$, different 
from the Laughlin sequence.
The QE correlation energy $u$ was calculated from the same 
Eq.~(\ref{eq1}), but with a different $\zeta=\sqrt{Q(Q+1)^{-1}}$ 
(where also $g=2Q+1$).
By using different $\zeta_{\rm QER}$ and $\zeta_{\rm QE}$ we removed 
discrepancy between $\lambda/R$ of finite $N$-QE$_{\rm R}$ and 
$N$-QE systems, in order to improve size convergence of $\Delta u=
u_{\rm QE}-u_{\rm QER}$.

In an ideal 2D system ($w=0$), the extrapolated value at $N^{-1}=0$ 
is $u_{\rm QE}=-0.013\,e^2/\lambda$, twice smaller (in the absolute 
value) than $u_{\rm QER}$ of a Laughlin state.
The difference $\Delta u=0.013\,e^2/\lambda$ is the key numerical 
result of this paper.
The accuracy of this estimate can be judged from the extrapolation
plot in Fig.~\ref{fig3}(b).

The fact that $u_{\rm QER}<u_{\rm QE}$ can be explained from the
comparison\cite{clusters} of QE$_{\rm R}$ and QE charge-density 
profiles $\rho(r)$.
The roughly gaussian $\rho_{\rm QER}$ is (up to normalization) very 
similar to $\rho_0$ of an electron in the lowest LL, yielding similar 
QE$_{\rm R}$ and electron pseudopotentials $V(\mathcal{R})$ and 
correlation energies $u$ (in the $\eta$-rescaled units).
The ring-like $\rho_{\rm QE}$ is more complicated and has a bigger 
radius, causing stronger (on the average) QE--QE repulsion.
The estimate of how much stronger -- depends on the accurate matching 
of the short- and long-range QE--QE pseudopotentials in Fig.~\ref{fig2}(a).
Therefore, to gain more confidence, we compared $u_{\rm QER}$ with 
$u_1$ of the electrons filling $\nu={1\over3}$ of LL$_1$, whose $\rho_1$ 
falls between $\rho_{\rm QER}\sim\rho_0$ and $\rho_{\rm QE}$ in terms 
of occupied area and the number of oscillations.
For the known\cite{fivehalf} $g=3N-6$ sequence of non-degenerate $\nu=
{1\over3}$ ground states in LL$_1$ we obtained $u_1=-0.32\,e^2/\lambda$.
Upon rescaling for the fractional QP charge, $\eta u_1=-0.021
\,e^2/\lambda$ falls between $\eta u_0\approx u_{\rm QER}=-0.026\,e^2/
\lambda$ and $u_{\rm QE}=-0.013$.
This demonstrates that the difference between $u_{\rm QER}$ and 
$u_{\rm QE}$ is caused by the difference between $\rho_{\rm QER}$ 
and $\rho_{\rm QE}$, and supports the obtained order of 
magnitude of $\Delta u$.

To demonstrate dependence of the correlation energies on layer width, 
in Figs.~\ref{fig3}(a) and (b) we also showed data for $w=8\lambda$.
The extrapolated values for this very wide layer are $u_{\rm QER}=
-0.025\,e^2/\lambda$ and $u_{\rm QE}=-0.031\,e^2/\lambda$.
Significant decrease of both energies compared to $w=0$ reflects an 
overall (averaged over in-plane distances, i.e., over $\mathcal{R}$) 
reduction of the QP repulsion in wider wells caused by the spread of 
electron (and thus also QE$_{\rm R}$ and QE) wavefunctions in the 
$z$-direction.
Due to different in-plane dynamics, $u_{\rm QER}$ and $u_{\rm QE}$ 
depend differently on width, and their difference $\Delta u=0.06\,
e^2/\lambda$ at $w=8\lambda$ is about twice smaller than at $w=0$.

\section{Spin phase diagram for $\nu_e=4/11$} 

Whether QEs or QE$_{\rm R}$'s will form a $\nu={1\over3}$ state at 
$\nu_e={4\over11}$ depends on the competition of Coulomb and Zeeman 
energies.
The condition for the QE$\leftrightarrow$QE$_{\rm R}$ transition is
\begin{equation}
   \Delta\varepsilon+\Delta u=E_{\rm Z}.
\label{eq2}
\end{equation}
The competing phases differ in electron spin polarization ($P=100$\% 
vs 50\%).
They are both incompressible, but probably have different excitation
gaps (and thus might not show equally strong FQH effect).
In an ideal 2D electron layer, the excitation gap (for neutral excitations) 
of the polarized state can be expected\cite{qepair} below $0.005\,e^2/\lambda$, 
and for the Laughlin state of QE$_{\rm R}$'s it is estimated at $\sim0.06
\eta\,e^2/\lambda=0.004\,e^2/\lambda$ (note, however, that a much smaller 
value $\sim0.001\,e^2/\lambda$ was predicted in Ref.~\onlinecite{Chang03}).
The nature of charged excitations, and the corresponding transport gaps
(especially in more realistic conditions, i.e., for $w>0$, including
LL mixing and disorder, etc.) are not known, and their prediction should
require a much more extensive calculation.

Let us concentrate on the question of stability of either QE$_{\rm R}$'s 
or QEs at $\nu_e={4\over11}$.
In order to draw in Fig.~\ref{fig3}(c) the phase diagram for GaAs 
heterostructures, we combined the estimated dependences of 
$\Delta\varepsilon/(e^2\lambda^{-1})$ and $\Delta u/(e^2\lambda^{-1})$ 
on $w/\lambda$ (where $e^2\lambda^{-1}/\sqrt{B}=4.49$~meV/T$^{1/2}$ 
and $\lambda\sqrt{B}=25.6$~nm\,T$^{1/2}$) with published data
\cite{Snelling91} on width dependence of the effective Land\'e factor 
$g^*$, governing the Zeeman splitting $E_{\rm Z}=g^*\mu_{\rm B}B$ 
(for $W\ge30$~nm, it is $g^*=-0.44$ and $E_{\rm Z}/B=0.03$~meV/T; in 
narrower wells, $g^*$ increases, passing through zero at $W\approx
5.5$~nm; recall that $w\approx W+3.3$~nm).

The most important phase boundary drawn in Fig.~\ref{fig3}(c) divides
the polarized and unpolarized $\nu_e={4\over11}$ states, i.e., the 
correlated QE and QE$_{\rm R}$ liquids at a finite $\nu={1\over3}$.
In experiment of Pan {\em et al.}\cite{Pan03} the polarized $\nu_e=
{4\over11}$ state was observed in a symmetric $W=50$~nm GaAs quantum 
well at $B=11$~T.
The corresponding point $(w,B)$ lies very close to predicted phase 
boundary, suggesting that the experimentally detected polarization 
depended critically on the choice of a very wide well.
It is clear from Fig.~\ref{fig3}(c) that the spin transition in narrower 
wells shifts quickly to higher magnetic fields (i.e., to higher electron 
concentrations $\varrho_e=\nu_e(2\pi\lambda^2)^{-1}$), especially when
the width dependence of $g^*$ is taken into account.
This suggests that the spin transition at $\nu_e={4\over11}$ might be 
confirmed in a similar experiment, carried out in a sample with the same 
$W$ and $\varrho_e$, but with the layer width $w$ tuned by the electric 
gates (inducing a controlled well asymmetry).

The role of QP interaction in stabilizing the QE$_{\rm R}$ phase is clear 
from the comparison of boundaries dividing correlated QE/QE$_{\rm R}$ 
liquids and non-interacting QE/QE$_{\rm R}$ gases
(the gas occurs at $\nu\ll{1\over3}$, with the critical equation $\Delta
\varepsilon=E_{\rm Z}$; the CF gas$\leftrightarrow$liquid transition was 
recently demonstrated by inelastic light scattering\cite{Gallais06}).
Additional boundaries (not shown here, but cf.\ Fig.~13(b) in 
Ref.~\onlinecite{sky}) appear at even smaller $B$, defining the areas 
of stability for a gas of CF skyrmions of different sizes.
\cite{Kamilla96,MacDonald98,sky,Leadley97}
Note also that $\Delta\varepsilon$ is determined more accurately than 
$\Delta u$, possibly explaining the incorrect position of the experimental 
point inside the predicted QE-gas/QE$_{\rm R}$-liquid area.

\section{Conclusion}

Combining composite fermion theory with exact numerical diagonalization 
we studied two spin states of the ``second-generation'' incompressible 
quantum liquid at $\nu_e={4\over11}$.
Our main result is prediction of a transition between these competing 
states, different not only by the spin polarization, but also by the 
microscopic mechanism of incompressibility (the nature of CF--CF 
correlation).
Starting with effective interaction pseudopotentials of polarized
and reversed-spin Laughlin quasielectrons (QE and QE$_{\rm R}$), we 
determined their correlation energies $u$ in conditions adequate for
realistic 2D electron layers of different widths $w$ and in different 
magnetic fields $B$.
This allowed us to draw a spin phase diagram of the $\nu_e={4\over11}$ 
state in the $(w,B)$ coordinates.
Comparison of our numerics with the experiment of Pan {\em et al.} 
is not conclusive.
However, our prediction of the spin transition induced in the same 
quantum well by external electric gates offers a possibility of more
accurate testing of the theory.
Finally, we have only considered pure QE or QE$_{\rm R}$ states
(i.e., confined ourselves to the extreme polarizations of $P=100$\% 
and 50\% in constructing the $\nu_e={4\over11}$ phase diagram),
leaving out the possibility of mixed QE/QE$_{\rm R}$ states with 
intermediate $P$ near the predicted phase boundary.

\section*{Acknowledgment}

The authors thank Wei Pan for helpful discussions and sharing 
unpublished experimental data, and acknowledge partial support 
from grants DE-FG 02-97ER45657 of US DOE and N202-071-32/1513 
of the Polish MNiSW.

\end{document}